\documentstyle[psfig,epsf]{EuroPhys}
\input EuroMacr
\def\lesssim{\mathrel{\mathpalette\vereq<}}
\def\vereq#1#2{\lower3pt\vbox{\baselineskip1.5pt \lineskip1.5pt 
\ialign{$#1\hfill##\hfil$\crcr#2\crcr\sim\crcr}}}

\begin{document}
\euro{}{}{}{}
\Date{}
\shorttitle{A. T\"OLLE \etal WAVENUMBER DEPENDENCE OF ETC.}
\title{Wavenumber dependence of structural {\boldmath$\alpha$}
       relaxation in a molecular liquid}
\author{A.~T\"olle\inst{1}
\footnote{email:albert@fuj.physik.uni-dortmund.de}, 
        H.~Schober\inst{2},
        J.~Wuttke\inst{3} \And
        F.~Fujara\inst{1}
       }
\institute{
\inst{1} Fachbereich Physik, Universit\"at Dortmund, D--44221 Dortmund \\
\inst{2} Institute Laue--Langevin, F--38042 Grenoble, France \\
\inst{3} Physik--Department E13, TU--M\"unchen, D--85747 Garching
          }
\rec{}{}
\pacs{
\Pacs{61}{20Lc}{Relaxations, time dependent properties}
\Pacs{61}{25Em}{Molecular liquids}
\Pacs{64}{70Pf}{Liquid glass transition}
}
\maketitle
\begin{abstract}
Structural $\alpha$ relaxation in liquid orthoterphenyl is studied by 
means of coherent neutron time-of-flight and backscattering spectroscopy
over a large temperature range.
Not only amplitude and relaxation time but also the spectral
line shape show a significant variation with wavenumber.
As expected from mode coupling theory, these variations are correlated
with the static structure factor. 
Even far above the melting point, $\alpha$ relaxation remains
non-exponential.
\end{abstract}
The most outstanding features of structural $\alpha$ relaxation
in glass-forming liquids are a strong temperature dependence
of the relaxation time and the stretching phenomenon of the
response function \cite{MCT}.
Very often, the correlation function $\Phi_X(t)$ of some
observable $X$ obeys time temperature superposition (TTS): 
by a shift of the characteristic relaxation time $\tau_X(T)$ 
the response functions at different temperatures can be superimposed
$\Phi_X(t;T)=\hat \Phi_X(\hat t),$ with $\hat t=t/\tau_X(T).$
While the master function $\hat \Phi_X(\hat t)$ varies weakly with 
temperature, the scale $\tau_X(T)$ increases rapidly upon cooling.
The other remarkable feature is the stretching phenomenon: 
the decay of $\Phi_X$ extends 
over a wider time range than expected for exponential relaxation;
empirically, it is generally well described by the Kohlrausch 
function (KWW)
\begin{equation}
\Phi_X(t)=A_X \exp[-(t/\tau_X)^{\beta_X}], \quad \beta_X<1.
\end{equation}\label{kww}

Different observables~$X$ may have not only different relaxation 
times~$\tau_X$ but also different stretching exponents~$\beta_X$.
In light scattering, for instance,
$\beta$ can be changed by just switching between polarized 
and depolarized geometry.
Uncountable experiments have confirmed both observations
for times varying from nanoseconds to hours, days, or longer.

A microscopic explanation of stretched relaxation has been given by
mode coupling theory (MCT) \cite{x15}.
Within MCT, time correlations decay in a two-step process which becomes
critically slow upon cooling due to a nonlinear feed--back mechanism 
in the microscopic dynamics of the particles.
A fractal time dependence in the late intermediate time
regime ($\beta$ relaxation) $\Phi_X(t)=f_X-h_X c(T) t^{-b},\, b>0$
leads to the final decay of correlations and is the key to stretching 
\cite{Got85}. 
However, it does not necessarily match the short-time expansion of eq.~1;
therefore, stretching exponents $\beta_X$ must not equal~$b$ \cite{x16}.
As the feed--back effect is expressed by a memory function which depends 
on the static structure factor $S(Q)$ the parameters for the density 
correlator become wavenumber ($Q$) dependent.
Numerical solutions of MCT equations for hard spheres \cite{FuHL92} and 
a binary mixture of soft spheres \cite{FuLa93} show pronounced oscillations 
of the parameters $A_Q$, $\tau_Q$ and $\beta_Q$ (eq.~1) in phase with $S(Q)$.

Experiments on molecular, ionic and even polymeric liquids
have confirmed the predicted trend of $A_Q$ and $\tau_Q$
to oscillate in phase with the static structure $S(Q)$ \cite{PeWu95}.
The wavenumber dependence of $\alpha$ line shapes has been studied far less.
For intensity reasons, high-resolution neutron scattering experiments
have most often been restricted to the structure factor maximum \cite{PeWu95}.
Only in polybutadiene, a spin-echo measurement has been performed in the
valley between the two maxima of $S(Q)$ 
where the $\alpha$ relaxation was found to be
more stretched than at the first maximum \cite{FrFR90}.
However, the oscillations of $\tau_Q$ get lost and $\beta_Q$ becomes
$Q$--independent when intra-chain motions are considered 
\cite{ArBW97}.
Therefore, experimental evidence of the predicted oscillations
of $\beta_Q$ is again missing.
For supercooled water, molecular dynamics indicates
weak oscillations of the stretching exponent~$\beta_Q$ in phase with~$S(Q)$
\cite{ScFC97}.

In this Letter we present a neutron scattering study 
of the $\alpha$ scale and the stretching in a molecular liquid
for different observables which are the closest related: 
density fluctuations for different wave vectors $Q$.
Particular emphasis is given on the wavenumber dependence.
Neutron scattering is a particularly unique tool since it couples 
directly to density fluctuations and covers the 
microscopic time range as well as the length scale of the 
intermolecular correlations.
While incoherent scattering from a protonated sample
measures the tagged particle motion,
collective motion is probed through the 
dynamic structure factor $S(Q,\omega)$
in coherent scattering from a deuterated sample.
Any wavenumber dependence of the coherent dynamic structure factor
may help to discriminate between different theories of $\alpha$ relaxation.

The experiments were performed 
on the time--of--f\/light spectrometer (TOF) IN5 
and on the backscattering spectrometer (BS) IN16 
of the Institute Laue Langevin in Grenoble, France.
The incident wavelengths were set to 6.5\,\AA$^{-1}$ 
and 6.27\,\AA$^{-1}$
leading to resolutions (fwhm) of $\sim$25\,$\mu$eV 
and $\sim$1\,$\mu$eV, respectively,
which allows us to detect relaxation processes from the psec (TOF) to
nsec (BS) time range.
In the BS experiment part of the analyser surface was covered with 
large Cadmium shields in order to increase the $Q$ resolution.
Vanadium was used to calibrate the detectors and yielded
the resolution functions.
Our system is the widely studied molecular liquid orthoterphenyl 
(OTP: C$_{18}$H$_{14}$, $T_{\rm m}=329$\,K, $T_{\rm g}=243$\,K).
A thin walled hollow aluminum cylinder filled with 99\% deuterated
OTP \cite{ToSW97} was used to keep
self absorption and parasite scattering as isotropic as possible.
The TOF--data were interpolated to constant $Q$ with step 
$\Delta Q=0.05\,${\AA}$^{-1}$. 
After Fourier deconvolution with the measured resolution function
and division by the static structure factor $S(Q)$ we obtain the
density correlator $\Phi_Q(t)=S(Q,t)/S(Q)$.

Fig.~1a presents the density correlators (IN5) 
at the structure factor maximum $Q_{\rm p}=1.45$\,\AA$^{-1}$ 
for temperatures between 313 and 400\,K.
In agreement with MCT all correlators show a two--step decay.
The second decay ($\alpha$ relaxation) becomes slower and more separated
with decreasing temperature and exhibits a pronounced stretching.
Fig.~1b shows the correlator at 360\,K for four $Q$ values.
We recognize that the $\alpha$ relaxation time 
scale on which the correlator decays to zero 
as well as its shape exhibit a rather non trivial $Q$ dependence.
The relaxation time $\tau_Q$ does not follow a simple power law 
$Q^{-2}$ expected for diffusion like motion.
At $Q_{\rm p}=1.45$\,\AA$^{-1}$ the correlations decay steeper
{\it i.~e.}~with {\it less} stretching than for other wave numbers.
This observation is also valid for the other temperatures where the
$\alpha$ relaxation falls into the accessible time window.
\begin{figure}[!t]
\epsfxsize=140mm
\centerline{\epsffile[34 553 536 754]{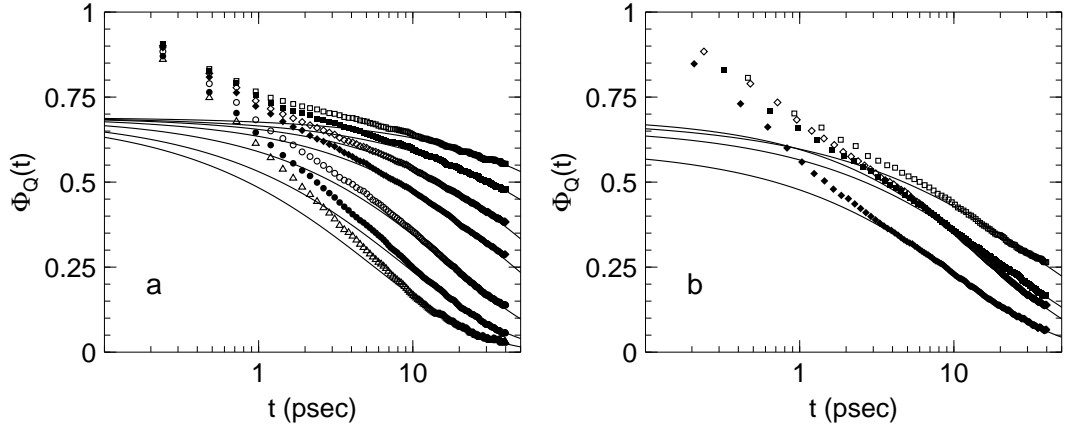}}
\caption
{(a) Correlators as measured on IN5 at $Q_{\rm p}=1.45$\,\AA$^{-1}$ 
for (from top to bottom) 313,\,320,\,330,\,340,\,360,\,380 
and 400\,K.
Solid lines are fits with eq.~1 with the parameters as determined from
the master curve discussed in the text.
(b) $\Phi_Q(t)$ at 360\, K for $Q$ values:
$Q$=0.95 (squares),\,1.2 (full squares),
\,1.45 (lozenge),\,1.7 (full lozenge)\,\AA$^{-1}$.
Note, at $Q_{\rm p}=1.45$\,\AA$^{-1}$ $\Phi_Q(t)$ decays steeper than the
other $Q$ values.}
\end{figure}

The spectral range of one instrument hardly allows to determine
the line shape of a complex relaxation from a single measurement.   
However, consistent results with acceptable error bars can be 
obtained by combination of measurements from different temperatures.

In our earlier work on OTP we found for selected $Q$ that
the relaxation time scales with viscosity  
$\tau_Q \propto \eta(T)/T$ and that the line shape is temperature 
independent \cite{old}.
Here, preliminary fits to the long time tail of the correlators
indicate that this viscosity scaling holds equally for all $Q$
over the whole temperature range.
Using published viscosity data \cite{viscosity}
we construct master curves $\Phi_Q(\hat t)$ via 
$\hat t = t[\eta(T_0)/T_0]/[\eta(T)/T]$
with an arbitrary normalization $T_0=290$\,K.
From a KWW fit to $\Phi_Q(\hat t)$ a first approximation of
$A_Q$, $\beta_Q$ and $\tau_Q(T_0)$ is obtained.
Fixing $\beta_Q$ and $\tau_Q(T)$ in fits to the individual $\Phi_Q(t)$
results in a set of amplitudes $A_Q(T)$ which are then used
to calculate reduced correlators $\Phi_Q(\hat t)/A_Q$.
This procedure is repeated and converges after 1 (3) iteration(s) 
for the TOF (BS) data.
The amplitudes $A_Q$ turn our to be temperature independent 
(as can be seen directly from the successful first rescaling,
cf.~fig.~2)
in agreement with a recent light scattering result in OTP in
the same temperature range \cite{CuLD97}.

From fig.~2 and~3 we recognize that the TTS works quite well.
(Note, from the TOF data we disregard the fraction of the correlators
which do not scale {\it e.~g.}~the microscopic dynamics and the 
$\beta$ process; 
in the BS experiment we observe only $\alpha$ relaxation.).
Both data sets show that the density correlations
around $Q_{\rm p}$ decay with less stretching and the relaxation
time is slightly slowed down compared to a $Q^{-2}$ dependence.
In order to quantify these effects eq.~1 is fitted to the long time tail 
of the master curves.
\begin{figure}[!t]
\epsfxsize=140mm
\centerline{\epsffile[34 374 536 754]{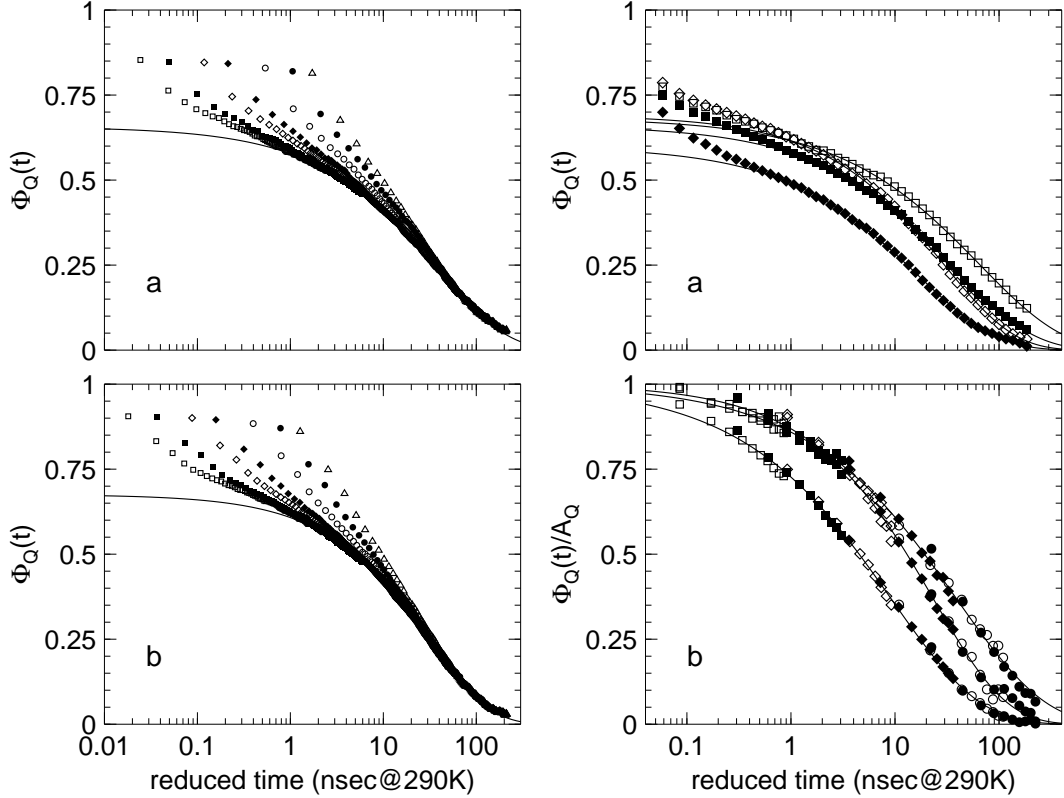}}
\caption{Master curves $\Phi_Q(\hat t)$ constructed via viscosity
scaling from TOF--data for $Q$=1.2\, \AA$^{-1}$ (a) and 
$Q$=1.45\, \AA$^{-1}$ (b).
Results from the first rescaling the times are shown 
without any adjustment of the amplitudes.
Full lines are fits with eq.~1 for reduced times $\hat t$ longer 
than 2 nsec.}
\end{figure}
\begin{figure}[ht]
\caption{(a) Master curve $\Phi_Q(\hat t)$ from TOF data 
for the same $Q$ values as in fig.~1b. 
(b) Normalized master curve $\Phi_Q(\hat t)/A_Q$ form BS data for three 
values of $Q$ obtained after the rescaling procedure explained in the text 
for 288, 293, 298, 305, 313 and 320\, K.
Both experiments show: at $Q_{\rm p}$ the relaxation is less stretched and 
the relaxation time is slowed down compared to a $Q^{-2}$ dependence.
}
\end{figure}

\begin{figure}[!t]
\epsfxsize=140mm
\centerline{\epsffile[80 350 508 753]{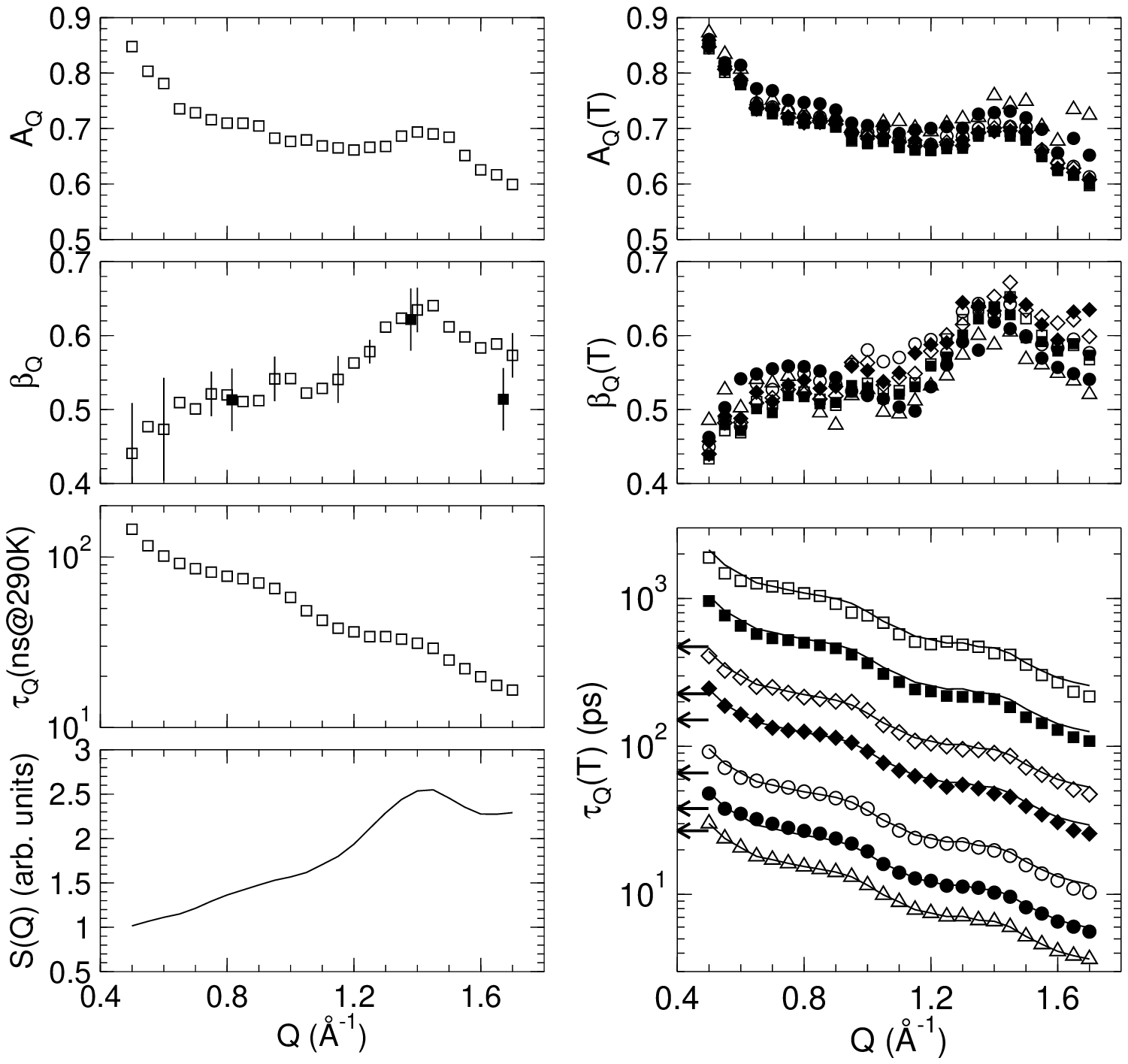}}
\caption{Left side: Fitting parameters of the master curve according to the
stretched exponential function eq.~1.
Note the presence of oscillations in all three parameters.
Errors of $A_Q$ and $\tau_Q$ are of the size of the symbol; error 
in $\beta_Q$ results form temperature uncertainty.
Right side: Kohlrausch parameters as determined from fits to the individual 
correlators in which the other two parameters were kept fixed to the values
obtained from the scaling analysis. 
Amplitude $A_Q$ and stretching $\beta_Q$ are roughly temperature independent.
The relaxation times $\tau_Q$ are compared to $\eta(T)/T$ which is 
modulated by the generic $Q$ dependence and scaled properly. 
Symbols are the same as in fig.~1a.
The arrows indicate results from depolarized light scattering ($Q\simeq0$) 
{\protect\cite{CuLD97}} for 320, 330, 340, 360, 380, 395\,K.
}
\end{figure}

In fig.~4 the fit results are compared to the static structure factor 
$S(Q)$ at 380\,K obtained by integration over the kinematically 
accessible region of IN5.
We note the presence of oscillations in all parameters 
highly correlated with the oscillations in $S(Q)$ at 
the first diffraction peak \cite{ToSW97,Mas97} 
around $Q_{0} \simeq 0.9$\,\AA$^{-1}$ and 
first structure factor maximum at $Q_{\rm p} \simeq 1.45$\,\AA$^{-1}$.
$A_Q$ is in qualitative agreement with the Debye--Waller factor $f_Q$ 
obtained from the analysis of the fast $\beta$ relaxation \cite{ToSW97}.
Both, the BS and the TOF experiment yield independently a maximum of
the exponent parameter $\beta_Q$ at $Q_{\rm p}$ whose
value is compatible with the one obtained earlier in a spin--echo 
experiment \cite{old}. 
No sign for a further oscillation in $\beta_Q$ is found.
The reduced relaxation time $\tau_Q$ shows two plateaus
around $Q_{0}$ and $Q_{\rm p}$
modulating an overall $Q^{-2}$ dependence.
We would like to point out that $\tau_Q$ and $\beta_Q$ depend on the 
rescaled data of a rather narrow temperature range 
{\it i.~e.}~340 - 380\,K for TOF and 298 - 313\,K for BS.
A possible slight temperature dependence of the parameters 
will not alter the results.
The absolute values depend on the quality of scaling;
their $Q$ dependence, however, is not affected.
Notwithstanding, the variations of $\tau_Q$ and $\beta_Q$ with $Q$
are clearly present in the raw data (fig~1,2,3).
At very low wave numbers $(Q\lesssim0.6$\,\AA$^{-1}$), where
$A_Q$ tends towards 1 the results may be spoiled by 
the remaining 15\% incoherent scattering which contribute particularly 
strong to the dynamic structure factor where $S(Q)$ itself is small.
A separation of coherent and incoherent scattering by the use of 
polarized neutrons would be helpful.

In order to get the temperature dependence of the parameters we
fit eq.~1 to the individual correlators fixing mutually the two others
to the values obtained from the analysis of the master curve.
Fig.~4 shows that the viscosity scaling is obeyed 
within experimental accuracy 
over several decades for all wavenumbers $Q$ 
(This corroborates the observation that
the quality of scaling is nearly identical for all $Q$, cf.~fig.~2.).
The curves corresponding to different temperatures are only shifted 
vertically due to the temperature dependence of the relaxation time
just what is expected if $\alpha$--scale universality holds.
The $Q$ dependence is given by a temperature independent function $C(Q)$.
Note the appearance of the two plateaus for all temperatures.
Results from depolarized light scattering($Q\simeq0$) 
\cite{CuLD97} are included for comparison. 
Finally, $A_Q$  and $\beta_Q$ are in a good approximation independent of 
temperature as observed earlier \cite{old,CuLD97}. 
For all temperatures qualitatively the same oscillations are found 
so they are clearly significant.
No indications for an increase of $\beta_Q(T)$ with increasing temperature
towards $\beta=1$ is found as in ref.~\cite{CuLD97}.
Even 70\, K above the melting temperature where one would na\"\i vely 
expect ordinary diffusion a reminder of a two step decay is still seen.

In conclusion, 
the $\alpha$ relaxation time for density fluctuations $\tau_Q(T)$ obeys
$\alpha$--scale universality;
amplitude $A_Q$ and stretching parameter $\beta_Q$ are 
in good approximation temperature independent.
The $Q$ dependence of amplitude $A_Q$, relaxation time $\tau_Q$ and 
stretching parameter $\beta_Q$ 
are qualitatively similar to the one predicted 
by MCT 
all oscillating in phase with the static structure factor $S(Q)$.
Correlations which are preferred by the system and build up the static 
structure factor maxima decay more slowly and with less stretching.

{\bf Acknowledgments}
We gratefully acknowledge stimulating discussions with 
W.~Kob, A.~Latz and H.~Sillescu.
We thank H.~Zimmermann (Max-Planck-Institut f\"ur medizinische Forschung, 
Heidelberg) for the deuteration of OTP and A.~Doerk for the help during sample 
preparation and H.~Z.~Cummins for providing the light scattering data.
Financial support of BMBF under project number 03--FU4DOR--4 is
appreciated.

\stars


\end{document}